\documentclass{aa}
\usepackage{graphicx}
\usepackage{txfonts} %
\usepackage{aalongtable}

\footnotesize
\newdimen\digitwidth    
\setbox0=\hbox{\rm0} \digitwidth=\wd0 \catcode`!=\active
\def!{\kern\digitwidth}
\normalsize

\begin{document}

\title{The Galactic distribution of magnetic fields in molecular clouds and HII regions}

\titlerunning{Magnetic fields in molecular clouds and HII regions}

\author
    {J.~L. Han
      \inst{1}
      \and J.~S. Zhang
      \inst{2,3}
    }

\institute
    {      National Astronomical Observatories, Chinese Academy of
           Sciences, Jia-20 DaTun Road, ChaoYang District, Beijing 100012, China
      \and Center for Astrophysics, Guangzhou University, 
           Guangzhou 510400, China
      \and Purple Mountain Observatory, Chinese Academy of
           Sciences, Nanjing 210008, China
    }
\date{Received 11 June 2006; accepted 27 October 2006. \\
{\bf Published as A\&A 464, 609-614 (2007)}. \\
http://dx.doi.org/10.1051/0004-6361:20065801.  Some typo fixed here.}

\abstract
{}
{Magnetic fields exist on all scales in our Galaxy. There is a controversy
about whether the magnetic fields in molecular clouds are preserved from the
permeated magnetic fields in the interstellar medium (ISM) during cloud
formation. We investigate this controversy using available data in the light
of the newly revealed magnetic field structure of the Galactic disk obtained
from pulsar rotation measures (RMs).}
{We collected measurements of the magnetic fields in molecular clouds,
including Zeeman splitting data of OH masers in clouds and OH or HI
absorption or emission lines of clouds themselves.  }
{The Zeeman data show structures in the sign distribution of the
line-of-sight component of the magnetic field. Compared to the large-scale
Galactic magnetic fields derived from pulsar RMs, the sign distribution of
the Zeeman data shows similar large-scale field reversals. Previous such
examinations were flawed in the over-simplified global model used for the
large-scale magnetic fields in the Galactic disk.}
{We conclude that the magnetic fields in the clouds may still "remember" the
directions of magnetic fields in the Galactic ISM to some extent, and could
be used as complementary tracers of the large-scale magnetic structure. More
Zeeman data of OH masers in widely distributed clouds are required.}

\keywords{ISM: molecules ­- ISM: magnetic fields ­- masers ­- magnetic fields ­-
  Galaxy: structure}

\maketitle

\section{Introduction}
The interstellar space is filled with HI gas, with a density of $n(H)$ of
about 1~cm$^{-3}$. The interstellar medium is not uniformly distributed;
rather, it is clumped. In some regions, the density is very high and
partially ionized, appearing as clouds with a size of several pc and a
density of $n(H) \sim 10^2$ to $10^3$~cm$^{-3}$. There are cores in some
clouds, where the density can be as high as 10$^{7}$cm$^{-3}$ and stars can
be formed there.

Magnetic fields permeate the interstellar medium as well as the clouds.  In
some clouds, magnetic fields are the dominant force against collapse by
self-gravity. Such clouds have been observed to have an hourglass
morphology, which indicates the strong regular magnetic field near the core
(see review of Heiles \& Crutcher 2005). The strength of magnetic fields in
clouds $|B|$ scales with the density $\rho$ as $|B|\propto\rho^{\sim0.5}$
(Crutcher 1999). The magnetic fields in clouds may be preserved when the
clouds were formed by contraction of diffuse interstellar medium. The
question then rises as to: whether or not the magnetic fields in molecular
clouds can still "remember" the large-scale magnetic fields in the
interstellar medium. Are they sufficiently strong that their correlation
with the large-scale fields was not destroyed by turbulence in clouds? If
so, the clouds can be an indepen- dent approach to reveal the large-scale
structure of Galactic magnetic fields. Conclusions from previous researches
on this subject (Davies 1974; Reid \& Silverstein 1990; Baudry et al.  1997;
Fish et al. 2003a) are contradictory.

     Magnetic fields in molecular clouds have been detected through
observations of Zeeman splitting of spectral lines for the line-of-sight
strength, and through polarized thermal emission from dust at mm, sub-mm or
infrared wavelengths for the transverse orientation of the fields. It is
difficult to observe the magnetic fields in diffuse interstellar medium and
molecular clouds (Heiles \& Crutcher 2005), because of the weakness of
fields and difficulties of calibration. Masers in massive star formation
regions near high density cloud cores, with a scale size of 100 AU, are very
bright (often $>$1 Jy) and their Zeeman splitting is relatively easier to
measure since the magnetic fields in such a dense region are very strong,
typically a few milligauss (mG).

     From Zeeman splitting data of a small sample of eight OH masers in the
ultra-compact HII regions excited by the central OB stars as well as of
seven HI clouds, Davies (1974) first noticed that the line-of-sight
direction of magnetic fields in all these clouds is parallel to the
direction of Galactic rotation, i.e. clockwise when viewed from the north
Galactic pole. Reid \& Silverstein (1990) pursued the idea and examined
published data available at that time, and obtained a sample of 17 reliable
OH maser sources with detectable Zeeman pairs. They noticed that 14 of 17
sources have line-of-sight magnetic field directions coincident with the
direction of the Galactic rotation, confirming the result of Davies
(1974). This surprising result is consistent with the magnetic fields in the
local arm region determined by pulsar RMs at that time (Manchester
1974). The implication is that the magnetic fields in the molecular clouds
are preserved during contraction from interstellar medium to star formation
region in clouds. Caswell \& Vaile (1995) observed 17 clouds with magnetic
fields in the the direction of Galactic rotation but also 11 in the
counter-rotation direction. Baudry et al. (1997) used their observations of
14 OH sources, together with 32 from the literature, and showed that 28 of
46 sources have fields in the direction of Galactic rotation (clockwise) and
the other 17 sources in the counterclockwise direction, excluding one near
the Galactic center. Fish et al. (2003a) identified 45 sources from survey
observations of massive star-formation regions, plus 29 sources in the
literature. They found that 41 of 74 sources are consistent with magnetic
fields in a clockwise sense, and 33 with fields in a counterclockwise
sense. This gives the impression that the maser data cannot be used to
reveal the global field structure of our Galaxy.

 The large-scale magnetic structure in the Galactic disk does not have one
dominant sense as originally hypothesized by Davies (1974). It has many
reversals. Recently Han et al. (2006) measured a large sample of pulsar RMs
and presented clear evidence for large-scale counterclockwise fields in the
spiral arms interior to the Sun and weaker evidence for a counterclockwise
field in the Perseus arm. In interarm regions, including the Solar
neighborhood, the evidence suggests that large-scale fields are clockwise.

    In this paper we collect all measurements of magnetic fields in star
formation regions as well as in molecular clouds, and compare them with the
magnetic field configuration newly derived from pulsar RM data. The
kinematic distances of molecular clouds or star formations regions are
unified to the frame of $R_0=8.5$~kpc and $V_{\odot}=220$km~s$^{-1}$. All
magnetic field measurements from Zeeman splitting observations are for the
line-of-sight component.

\section{Collected Zeeman splitting data for cloud magnetic fields}

Two methods have been used to measure the magnetic fields of molecular
clouds. The classical one is to observe the Zeeman splitting of HI or OH
lines (see below), which gives the strength of the magnetic field along the
line of sight. The second one is to map the polarization of clouds at mm,
submm or infrared (e.g.  Chuss et al. 2003), which can show the magnetic
field orientation in the sky-plane. However, there are not many molecular
clouds mapped by polarization. Therefore we will only collect the Zeeman
splitting data.

    Zeeman splitting of spectral lines occurs in two kinds of regions. OH or
other emission lines are very strong from maser spots in high density HII
regions or star formation regions, mainly in the ionized surrounding of
newly formed stars in molecular clouds. There is much data for this kind of
observation. The emission or absorption lines (OH or HI) are also observed
for nearer layers of clouds. We surveyed the literature for both types of
observations.

\subsection{Zeeman splitting observations of masers}

In the core of molecular clouds (often a star-formation region or HII
region), maser spots have been observed in the shock-front or
ionization-front surrounding newly formed stars (e.g. Reid et al. 1980;
Zheng et al. 2000). We collected all measurements of Zeeman splitting of OH
masers (10$^5 \sim 10^8$~cm$^{-3}$), but not H$_2$O masers for higher
density regions ($\sim10^{10}$~cm$^{-3}$). These data give the line-of-sight
direction of magnetic fields in situ at the location of masers. One cloud
often has many maser spots, and they show different field strengths and
sometimes different field directions, but always indicating an organized
magnetic field structure (e.g. Fish et al. 2005a; Fish \& Reid 2006;
Bartkiewicz et al. 2005).  

Masers have been observed with different
resolutions using single-dish telescopes (e.g. Parkes, Effelsberg) or
interferometers (e.g. VLA or ATCA, EVN or VLBA). It is important to have
high resolution observations to identify maser spots and derive the magnetic
fields in each of them. On the other hand, estimates for the magnetic fields
from single dish observations of masers can still be useful and meaningful
if there is no confusion (simple patterns) in the spectra of
masers. Although a low-resolution telescope beam would significantly reduce
the Zeeman splitting, being the intensity-weighted average even if all
masers have the same sign but are displaced in velocity (see Sarma et
al. 2001), the inferred field strength can be the mean value of the magnetic
fields at all maser spots (e.g. Caswell 2004a).

\begin{figure*}[bht]  
\begin{center}
\includegraphics[angle=270,width=130mm]{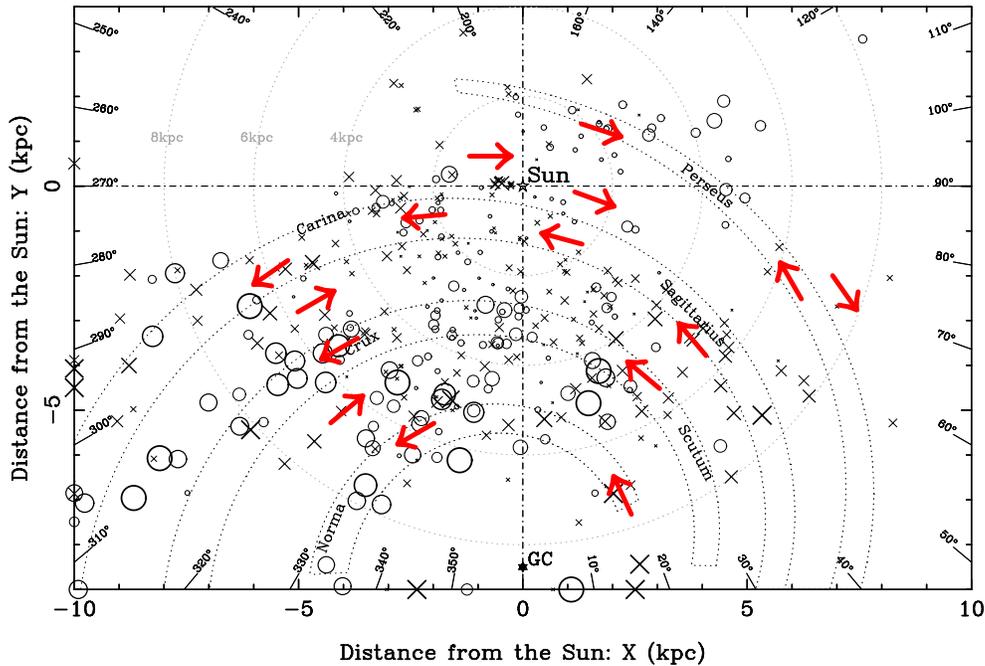}
\caption{The RM distribution of 374 pulsars with $|b|<8\degr$, projected
onto the Galactic Plane. The linear sizes of the symbols are proportional to
the square root of the RM values, with limits of 9 and 900 rad m$^{-2}$. The
crosses represent positive RMs (indicating the average field over path
pointing towards us), and the open circles represent negative RMs (for the
average fields going away from us). The approximate locations of four spiral
arms are indicated. The large-scale structure of magnetic fields derived
from pulsar RMs are indicated by thick arrows. See Han et al. (2006) for
details.}
\end{center}
\end{figure*}

    In Table 1, we list OH maser measurements collected from the
literature. We discarded the measurements with grade "D" in Fish et
al. (2003a) which are thought not to be reliable identifications as the
authors have claimed; and we also take the middle value if a field strength
range is given in their data. We give the two extremes as well as the median
value for magnetic field measurements in the table. If there is only one
measurement for a cloud, such as from the Parkes telescope by Caswell (see
references in the Table 1), or only for one maser spot from VLBI
observations, we give them in the median value. We sometimes discard old
measurements if there are too many new high resolution observations
available. Care should be taken to avoid misinterpretation of the data, such
as G285.26$-$0.05 at 1665 MHz in Davies (1974), see discussions in Caswell \&
Vaile (1995). The signs for field directions were sometimes not given in the
data tables (e.g. Gaume \& Mutel 1987; Baudry \& Diamond 1998; Desmurs et
al. 1998), so one has to look at the original plots of the RH/LH maser
spectrum, or text, or other references to identify their signs. In some
papers measurements were discussed in the text rather than expressed in data
tables (e.g. Caswell \& Vaile 1995). Distances of most of Caswell's table are
given with the old IAU standard (10 kpc/220 km s$^{-1}$ ) and should be
corrected to the current one (8.5 kpc/220 km s$^{-1}$). For ambiguous
kinematic distances we take the nearer one but marked it with an asterisk
after the value. We updated the distances of some HII regions according to
the latest reference (e.g. Fish et al. 2003b).  If one object has many
observations (e.g. of many lines), one good measurement was selected as
representative of the cloud.  For this purpose we normally take the best
result from high resolution spatial observation or the median of the many
median values, as marked with a "*".

    We did not use the Zeeman splitting data of OH masers associated with
proto-planetary nebulae, supernova remnants, or young stellar objects for
our study.

\subsection{Zeeman splitting observations of molecular clouds}
We searched the literature for measurements of magnetic fields in molecular
clouds. Crutcher (1999) has collected good measurements of 15 clouds in
Table 1 of his paper with detailed discussions on each cloud in the
appendix. We complemented this data with new measurements, and present them
in Table 2. The sources are given in the order of Galactic coordinates. The
relative information, such as distance, the emission or absorption line line
observed, the instruments as well as the frequency of observations are also
given. Observations of one object but by different authors or different
emission or absorption lines are listed in different lines in the
table. Note that different observation resolutions can cause different
results (see e.g. Brogan \& Troland 2001a), so we also list the telescopes
used for observations.  Similarly, we take the median of the many
observations or the measurement of highest quality as a representative of a
cloud.  For these we mark a "*" after the field strength.

We do not include the magnetic field measurements of diffuse clouds
(e.g. Myers et al. 1995; Goodman \& Heiles 1994) or HI filaments (Heiles
1989), which are not gravitationally bound and therefore are not molecular
clouds. Similarly, we did not include data of Zeeman splitting of absorption
lines from the cold neutral medium by observing extragalactic radio sources
(e.g.  Heiles \& Troland 2004).

\begin{figure*}[hbt]
\begin{center}
\includegraphics[angle=270,width=120mm]{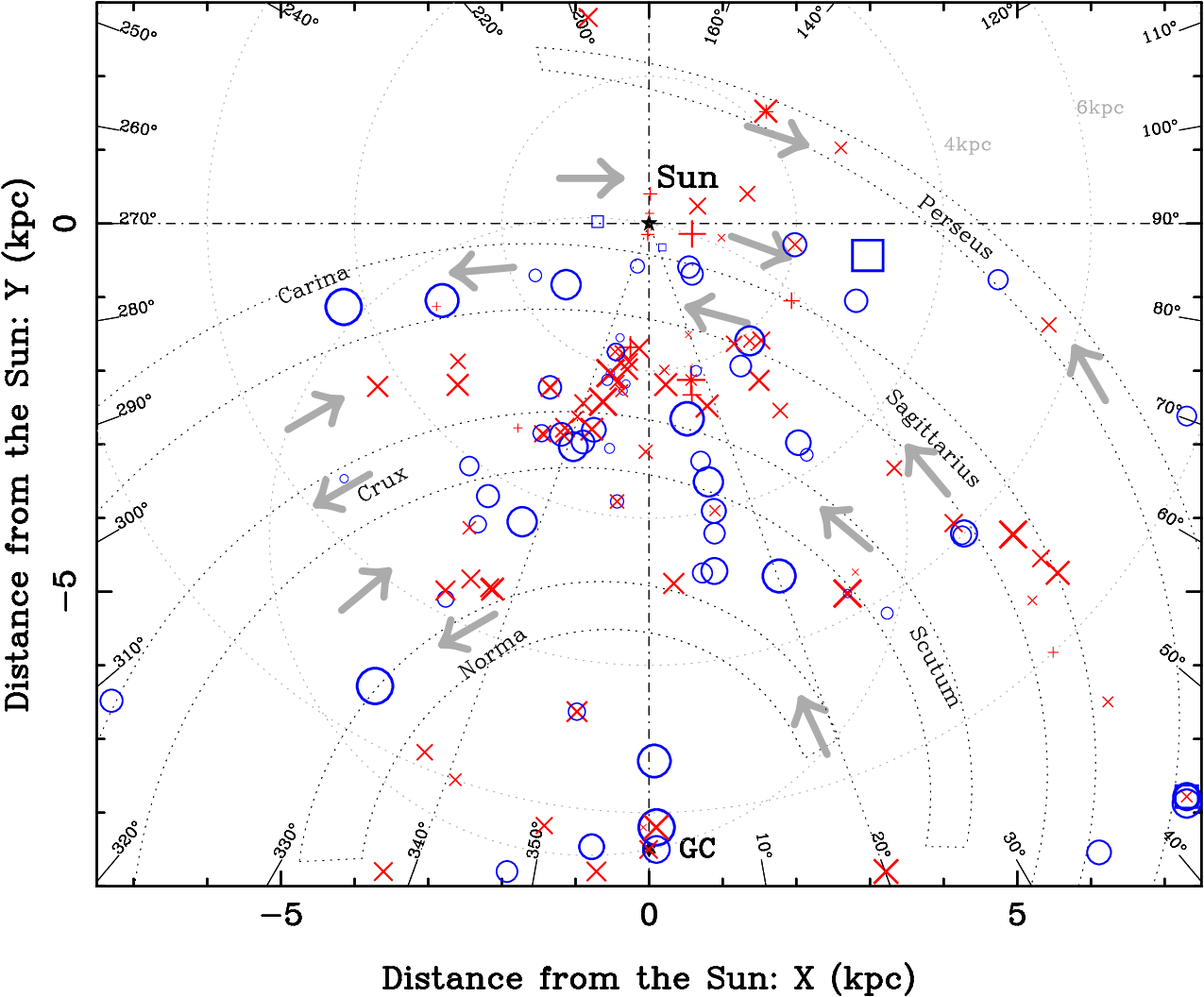}   
\caption{The medians of field measurements from Zeeman splitting of OH
masers (cross and circles) in 137 objects or HI or OH lines of 17 molecular
clouds (plus and squares) projected onto the Galactic plane, with the rough
indication of spiral arms and the magnetic field directions (arrows) derived
from pulsar RM data. The linear sizes of the symbols are proportional to the
square root of the field-strength values, with limits of 0.5 and 9 mG for
fields from the median maser fields and of 15 $\mu$G and 270 $\mu$G for
median cloud fields. The crosses or pluses on the right ($0\degr< l <
180\degr$ ) represent positive $B$, ie. the field direction going away from
observer, and circles or squares going towards us. The symbols on the left
($180\degr< l < 360\degr$) are reversed, so that all crosses and pluses are
consistent with the clockwise fields viewed from the Northern Galactic pole,
and all circles and squares with counterclockwise fields.
}
\end{center}
\end{figure*}

\section{Analysis and discussion}

The HII regions and molecular clouds are confined to the Galactic plane, and
are regarded as tracers of the spiral arms.  Here we extend the work by Fish
et al. (2003a), in light of the newly derived magnetic field configuration
associated with spiral arms in the Galactic disk from pulsar RMs (Han et
al. 2006) as well as more measurements of magnetic fields from masers and
molecular clouds.

\subsection{The Galactic magnetic fields derived from pulsar RMs}

The large-scale magnetic fields in the Galactic disk have been derived from
pulsar RMs (Han et al. 2006). The variation of RMs with the dispersion
measures of pulsars indicates the magnetic field direction (see arrows in
Fig. 1). There are many reversals of large-scale magnetic fields. In some
regions pulsar data were rich enough to derive the magnetic field direction.
However, the fields in many other regions cannot be determined due to
scarcity of data points. Also, random fields in some regions may be stronger
than the regular magnetic fields, which complicates derivation of field
directions from pulsar data. Our current knowledge of large-scale magnetic
fields is shown in Fig. 1, very different from the over-simplified field
structure used in previous studies (e.g. Davies 1974).

    At high Galactic latitudes, Han et al. (1997, 1999) identified a
striking antisymmetry in the RM distribution in the sky, mainly in the inner
Galaxy, which was argued as being caused by azimuthal magnetic fields in the
Galactic halo with reversed field directions below and above the Galactic
plane. However, there is little Zeeman splitting data for the molecular
clouds at high Galactic latitudes (see Tables 1 and 2 below), so we will not
discuss these data for the halo field. Instead, we concentrate on the
large-scale field in the Galactic disk.

\subsection{Overview of the data}

We plot the measurements of Zeeman splitting for magnetic fields in Fig. 2,
using the median values in Tables 1 and 2. The medians are good
representatives if many spots are observed with high resolution observations
for the Zeeman splitting in molecular clouds or star formation regions
(e.g. Fish \& Reid 2006; Fish et al. 2005b). If there are measurements for
only two spots, then we took their average. Given the fact that the
large-scale magnetic fields in the Galactic disk are not in one field sense
(clockwise) as assumed in many previous analyses, it is not meaningful to
count how many data points have the same sense as the Galactic rotation and
how many have the opposite sense. Rather, we should compare data with
proposed models for the magnetic fields in the Galactic disk.

    As seen in Fig. 2, the sign distribution of Zeeman splitting data have a
clear structure. In many regions there is a dominant direction of
line-of-sight components, which can be related to the clockwise (CW) or
counterclockwise (CCW) sense of the Galactic azimuthal magnetic fields. As
discussed by Fish et al. (2003a), most measurements (8 crosses or pluses of
10 data) outside the solar circle are consistent with a CW large-scale field
nearer than or around the Perseus arm. Second, most data points (circles) in
the Carina arm are consistent with a counterclockwise (CCW) large-scale
field derived from pulsar RMs. As noticed by Fish et al. (2003a), masers
(crosses) in the Sagittarius arm at distances farther than 6 kpc show a
coherent sense of CCW field direction that is in contrast to the large-scale
field from pulsar RMs. However the location of the arm has a large
uncertainty, very probably shifted inwards (see Cordes \& Lazio
2002). Third, between the Carina-Sagittarius arm and the Crux-Scutum arm,
Zeeman splitting data show the very dominant CW sense (crosses). Going
inwards, one can see that data (more circles) are dominantly consistent with
CCW large-scale fields in or near the Crux and Scutum arms. The data
(crosses) near the Norma arm show a reversed CW field, consistent with the
directions of the interarm field derived from pulsar RM data.  If the large
uncertainty of kinetic distances of molecular clouds or maser regions is
considered, such field reversals are similar to those newly identified by
Han et al. (2006) from pulsar RMs.

\subsection{Indication for large-scale field reversals?}

Fish et al. (2003a) suggested that the magnetic fields revealed by
masers are ordered or correlated on a scale of a few kpc. Here we tried to
check the sense-correlation of the data shown in Fig.2. If there is no
significant correlation, then the data do not contain information about
large-scale fields.

\begin{figure}
\begin{center}
\includegraphics[angle=270,width=88mm]{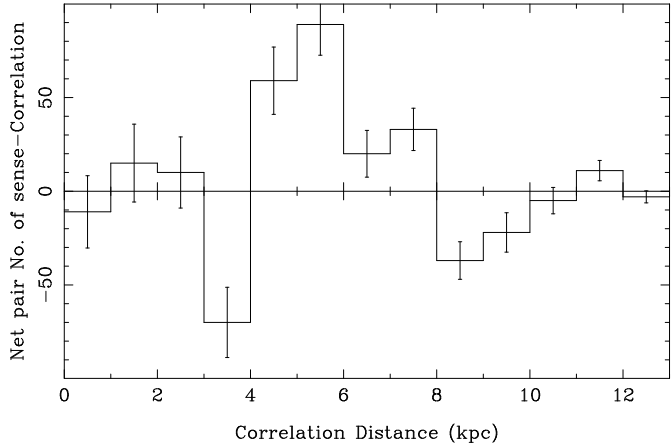}   
\caption{Correlation for senses of magnetic field data in Fig.2.
The error-bars were estimated by $\sqrt{N}/2$, here $N$ is the total
number of pairs in a bin. See text for details.}
\end{center}
\vspace{-3mm}
\end{figure}
We take $+1$ for all crosses or plus in Fig.~2, and $-1$ for all circles or
squares. Thus, a median field of maser region with a direction
consistent with CW sense is marked as $+1$, and that 
with CCW as $-$1. If a pair of regions at a given distance have
the same sense, they are correlated, and if they have the opposite
sense, they are anti-correlated. We consider the net correlation
pair numbers at different separation distances.

    The results are shown in Fig. 3: pairs of objects with a separation of
less than 3 kpc tend to a null correlation, due to either random fields or
ordered fields along the spiral arms with opposite senses. If they are
separated by 2 to 4 kpc, they tend to have the opposite sense. If they
separated by 4 to 8 kpc then they tend to have the same sense. Considering
the negative-positive oscillation of data probably due to sign-clusters
associated with different spiral arm or interarms, though with only marginal
significance, we believe that such a correlation is probably an indication
of large-scale field reversals in the Galactic disk.

\begin{figure}
\begin{center}
\includegraphics[angle=270,width=86mm]{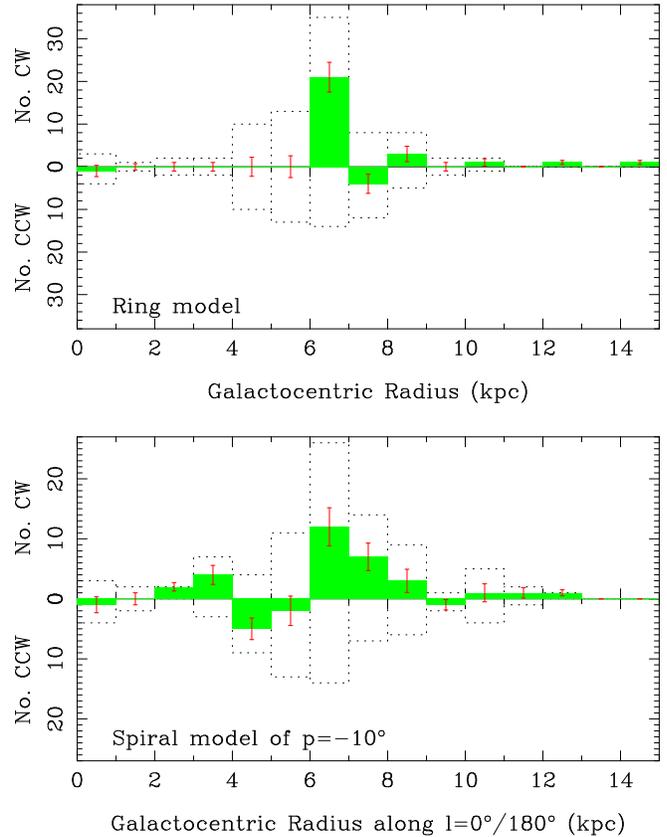}   
\caption{The histogram of the CW and CCW senses of median magnetic fields of
masers and clouds, binned for the different Galacto-centric radius
ranges. For the spiral model of a pitch angle of $-10\degr$, the
Galactocentric radii are scaled to that along $l=0\degr$ or $180\degr$.  The
gray area shows the net counts in a given Galactocentric radius, and
error-bars were estimated by $\sqrt{N}/2$, here $N$ is the total number of
data in a bin. Note that the field reversals have marginally been revealed by 
data in the frame of the spiral model.}
\end{center}
\vspace{-3mm}
\end{figure}

\subsection{Models for the Galactic magnetic fields and the Zeeman splitting data}

Now we can check how well the Zeeman data match the models for the
large-scale magnetic fields in the Galactic disk.

There are two models for magnetic fields in the Galactic disk, a
ring model in which the magnetic fields alternate their directions in many
concentric rings relative to the Galactic center (Rand \& Kulkarni 1989;
Rand \& Lyne 1994; Vall\'ee 2005), and a spiral model in which the magnetic
fields follow spiral arms but reverse their directions from arm to arm (Han
\& Qiao 1994; Indrani \& Deshpande 1999; Han et al. 2006). We took a pitch
angle of $p=-10\degr$ as the most probable value for the spiral.

As shown in Fig.4, for the concentric ring model, we count how many
maser regions and clouds show CW fields and how many show CCW in a
given range of Galactocentric radius $R_c$. If the counts are roughly
equal, this implies no dominant field direction in the radius range. The
net counts are shown in gray, indicating the dominant field directions in
those ranges. For the spiral model, assuming a pitch angle of $-10\degr$, we do
the same, but the Galactocentric radius of a maser region or cloud is
scaled to the Galactocentric radius $R_c$ along $l=0\degr$ or $180\degr$.

The dominant CW field in the range of $6<R_c<7$~kpc is clearly shown in the
ring model. The field sense is coincident with the Galactic rotation. In the
ring model, there may be no large-scale fields except in this radius
range. This is not true, as is shown in the magnetic fields in Fig.~1 and
the sense-correlation in Fig.2 at a distance greater than 2 or 3~kpc. The
counts in the spiral model showed the field reversals from the inner Galaxy
to the outer Galaxy, although some are only marginally significant, i.e.,
the CW fields at 2 to 4~kpc, the CCW at 4 to 6~kpc, the CW at 6 to 8~kpc and
the CCW about 9~kpc and the CW field outside 9~kpc.

The dominant CW data in the Sagittarius arm are not consistent with the
dominant CCW data in the Carina arm, so they diminished the net counts for
the arm. The field directions given by pulsar RM data, opposite to the
Zeeman data, are also similarly incompatible in the two arms. Nevertheless,
all data in the two arms independently show large-scale fields in the arms.

\subsection{Discussion}

There are still large uncertainties in this study. One is the uncertainty in
the determination of the large-scale magnetic field from pulsar RM
data. Although the data have been enriched in some regions, the large-scale
magnetic fields in many regions remain to be measured with more pulsar RM
data (Han et al. 2006) or extragalactic radio sources (Brown et
al. 2003). On the other hand, the Zeeman splitting data for {\it in situ}
measurements of magnetic fields in clouds have two problems. One is the
large uncertainty or ambiguity in the dynamic distances of clouds, which
could be 10\% or more (see G\'omez 2006). Another problem is how to relate
the field structure inside a cloud to the large-scale fields, which may be
better understood after more measurements of high resolution observations
become available (e.g. Fish et al. 2005a, Fish \& Reid 2006). We took the
median value in this study. Over many years, Caswell and Reid et al. have
made detections and measurements, so that much Zeeman splitting data have
been accumulated and have been included in this study. However, much more
Zeeman splitting data of masers and clouds, in wider regions and high
resolution observations, are needed.

Considering these uncertainties, we found that the reversals shown by data
of maser regions and clouds are similar to those of large-scale magnetic
fields derived from pulsar RMs. As suggested by previous authors (Reid \&
Silverstein 1990; Baudry et al. 1997), this may allow the use of the Zeeman
splitting measurements of magnetic fields in clouds to reveal the
large-scale magnetic fields in the our Galaxy.

Molecular clouds were formed by contraction of diffuse gas in the
interstellar medium, and the magnetic fields are so enhanced that they have
the same energy as the kinetic energy (Crutcher 1999). The role of magnetic
fields during such a contraction has been observed by the hourglass shape of
clouds, which is an indication of field direction conservation. Our
analysis above indicates that the field direction in clouds may be preserved
from the large-scale field in the ISM during the contraction.

How can such a coherence and consistence of magnetic field directions occur
from the low density of ISM ($n \sim 1 {\rm cm}^{-3}$) to higher density clouds
($\sim10^{3}{\rm cm}^{-3}$), even to the highest density maser regions
($\sim10^{7}{\rm cm}^{-3}$), after a density compression of about 3, or even
10, orders of magnitude? One implication of this result is that the clouds
probably do not rotate much after they are formed. Otherwise, the field
directions of clouds we measured would be random. During the process of star
formation, the clouds seem to be too heavy to be rotated, although there are
jets or disks from newly formed stars which may have some dynamic
effects. Furthermore, the fields in the molecular clouds are strong enough
after the contraction so that the turbulence in the clouds cannot
significantly alter the magnetic field status.

\section{Conclusion}

We have collected available Zeeman splitting measurements of magnetic fields
in molecular clouds from the literature, and found a sign-coherency of the
line-of-sight magnetic field component in many regions of the Galactic
disk. Such a sign distribution is closely related to spiral arms, and shows
similar field reversals to those newly derived from pulsar RM data. Thus,
the magnetic fields in molecular clouds may be related to the large-scale
field structure. If this can be confirmed by a larger dataset, the physical
picture of the contraction of cloud formation may be better understood, in
which both the density and magnetic fields are enhanced and the field
directions can be preserved.  The measurements of in situ magnetic fields in
molecular clouds could be used as probes for the Galactic-scale magnetic
field, complementary to pulsar RM data.

The molecular clouds and massive star formation regions where the masers are
observed are also tracers of spiral arms. To further test whether the
magnetic fields in molecular clouds are correlated with the large scale
Galactic field, Zeeman splitting data from more molecular clouds are needed
and their distance uncertainty or ambiguity should be resolved as well.

\begin{acknowledgements}
We thank Mr. Chen Wang for help and the anonymous referee and
Prof. R. Wielebinski and Prof. E. Falgarone for helpful comments on the
manuscript. JLH is supported by the National Natural Science Foundation of
China (10521001 and 10473015)
\end{acknowledgements}

{}





\begin{longtable}{lllrrrrll}
 \caption{\label{Binmasers} Magnetic field measurements from Zeeman splitting of OH masers}     \\
\hline\hline
Source &   Alias    &  !Dist   &  Freq  &   $B_l$ & $B_m$& $B_u$& Obs. & Ref.                   \\
(GL+GB)&   name     &  (kpc)  &  (MHz) &   (mG)      &  (mG)& (mG)   & telescope &              \\
\hline
\endfirsthead
\caption{continued.}\\
\hline\hline
Source &    Alias   &  !Dist  &  Freq   &   $B_l$ &  $B_m$&$B_u$& Obs. & Ref.                  \\
(GL+GB)&    name    &  (kpc) &  (MHz)  &  (mG)       &  (mG)   & (mG)    & telescope &         \\
\hline
\endhead
\hline
\endfoot

G000.547$-$0.852&RCW 142    &  !7.3 &  1667&       &$-$7.4* &      &   VLA-A & fram03                 \\
G000.666$-$0.029&...        &  !8.5 &  6035&       &$-$5.0* &      &   PKS   & c03                    \\
G000.666$-$0.034&Sgr B2M    &  !8.2 &  1720&       &   0.7! &      &   VLA-A & fram03                 \\
...             &...        &  !8.2 &  1720&    4.5&      * &  10.0&   VLA-C & gm87                   \\
G000.666$-$0.035&...        &  !8.2 &  1720&       &   9.!! &      &   PKS   & c04a                   \\
...             &Sgr B2N    &  !8.2 &  1665&       &   1.9! &      &   VLA-C & gm87                   \\
G003.910$+$0.001&...        &  !4.9 &  6035&       &   5.0* &      &   PKS   & c03                    \\
G005.885$-$0.392&...        &  !2.2 &  1720&       &   6.!* &      &   PKS   & c04a                   \\
\end{longtable}
\\
\multicolumn{9}{l}{\parbox{130mm}{
Notes:                                                                                   

1.     Bl \& Bu are the two extrema of magnetic field strengths of many
       maser spots in high resolution VLBI observations; we also give magnetic
       fields in Bm if there is only measurement for one spot. In low resolution single
       dish observations we give magnetic field strength in Bm, but if there are many B-
       values for different velocities, then we give two extrama for Bl
       and Bu as well as Bm. Signs are values extracted from the
       paper are markes with '*'.                    

2.     Distance: in kpc, (with 8.5/220 frame).
       If there is an ambiguity in kinematic distances, we take the nearer one
       and marked with '*'                                                                

3.     Positive B value indicates the field pointing away from us, and
       Negative towards us.      
}} \\
\end{longtable}

\begin{tabular}{llll}
\multicolumn{4}{l}{References:}  \\
bc85:   &Baart \& Cohen 1985, MNRAS 213, 641;                        &
bcd+86: &Baart et al. 1986, MNRAS 219, 145;			     \\
bfgt00: &Brogan et al. 2000, ApJ 537, 875;			     &
bgyr03: &Bains et al. 2003, MNRAS 338, 287;			     \\
brgy04: &Bains et al. 2004, MNRAS 354, 529;			     &
bscr05: &Bartkiewicz et al. 2005, MNRAS 361, 623;		     \\
bd98:   &Baudry \& Diamond 1998, A\&A 331, 697;			     &
bdwc97: &Baudry et al. 1997, A\&A 325, 255;			     \\
bmg84:  &Benson et al. 1984, AJ 89, 1391;			     &
c98:    &Caswell 1998, MNRAS 297, 215;				     \\
c01:    &Caswell 2001, MNRAS 326, 805;				     &
c03:    &Caswell 2003, MNRAS 341, 551;				     \\
c04a:   &Caswell 2004a, MNRAS 349, 99;				     &
c04b:   &Caswell 2004b, MNRAS 352, 101;				     \\
\end{tabular}


\begin{longtable}{lllrrrllll}
\caption{\label{Binclouds} Magnetic field measurements from Zeeman splitting of molecular clouds}                     \\
\hline\hline
 Source       &    Alias  & Dist &     B!!     &  $\sigma_B$
 &\multicolumn{1}{c}{abs./emi.} & !!Freq.& Obs. & Ref.  & Remarks \\
 (GL+GB)      &     name  &  (kpc)    &\multicolumn{1}{c}{($\mu$G)} & \multicolumn{1}{c}{($\mu$G)} & \multicolumn{1}{c}{line} & !!(MHz) & telescope &       &              \\                 
\hline
\endfirsthead
\caption{continued.}\\
\hline\hline
Source       &    Alias  &Dist   &     B!!      &  $\sigma_B$     &
\multicolumn{1}{c}{abs./emi.}  & !!Freq & Obs. & Ref.  & Remarks \\
(GL+GB)      &     name  &  (kpc)    &\multicolumn{1}{c}{($\mu$G)}& \multicolumn{1}{c}{($\mu$G)}  & \multicolumn{1}{c}{line} & !!(MHz)& telescope &       &          \\
\hline
\endhead
\hline
\endfoot

G000.70$-$00.04& SgrB2 NSW-Z    & !8.20 & $-$720! &160 &  absp-HI &  !!1420   & VLA-BA  & crmt96   &                \\
...            & SgrB2 Main-Z   & !8.20 & $-$480* &110 &  absp-HI &  !!1420   & VLA-BA  & crmt96   &  B-variable    \\
G014.00$-$00.60& G14.0$-$0.6    & !2.4  & !$+$64* & 18 &  absp-OH &  !!1665/7 & GB43    & bmrh01   &                \\
G015.00$-$00.70& M17SW (core)   & !2.2  & $-$450! &100 &  absp-HI &  !!1420   & VLA-DC  & btrc99   &  GMC-complex   \\
...            & M17NW          & !2.2  & $+$550! &100 &  absp-HI &  !!1420   & VLA-DC  & btrc99   &  B-structure   \\
G015.00$-$00.70& M17SW          & !2.2  & $+$150* & 50 &  absp-OH &  !!1665   & VLA-CB  & bt01b    &  GMC-complex   \\
...            & M17SW          & !2.2  & $+$260! & 45 &  absp-HI &  !!1420   & VLA-CB  & bt01b    &  B-structure   \\
G028.80$+$03.48& W40            & !0.37 & !$-$14* &  3 &  absp-OH &  !!1665/7 & Nancay  & ckt87    &  HII-MC        \\
\hline \hline
\end{longtable}

\begin{tabular}{llll}
\multicolumn{2}{l}{References:} \\
bmrh01: &Bourke et al. 2001, ApJ 554, 916;                &
bt01a:  &Brogan \& Troland 2001a, ApJ 550, 799;	  \\
bt01b:  &Brogan \& Troland 2001b, ApJ 560, 821;		  &
bta+05: &Brogan et al. 2005, ASP conf S. Vol.343, p.183;  \\
btrc99: &Brogan et al. 1999, ApJ 515, 304;		  &
c99:    &Crutcher 1999, ApJ 520, 706;			  \\
ck83:   &Crutcher \& Kaz\'es 1983, A\&A 125, L23;	  &
ckt87:  &Crutcher et al. 1987, A\&A 181, 119;		  \\
crmt96: &Crutcher et al. 1996, ApJ 462, L79;		  &
crtg99: &Crutcher et al. 1999a, ApJ 515, 275;		  \\
ct00:   &Crutcher \& Troland, 2000, ApJ 537, L139;	  &
ctg+93: &Crutcher et al. 1993, ApJ 407, 175; 		  \\
ctl+99: &Crutcher et al. 1999b, ApJ 514, L121;		  &
frm05:  &Fish et al. 2005b, ApJ 623, 269;		  \\
gh94:   &Goodman \& Heiles 1994, ApJ 424, 208;		  &
gch+89: &Goodman et al. 1989, ApJ 338, L61;		  \\
kc86:   &Kaz\'es \& Crutcher 1986, A\&A 164, 328;	  &
ktch88: &Kaz\'es et al. 1988, ApJ 335, 263;		  \\
mggh95: &Myers et al. 1995, ApJ 442, 177;		  &
strc00: &Sarma et al. 2000, ApJ 533, 271;                 \\
rctg93: &Roberts et al. 1993, ApJ 412, 675;		  & 
rct95:  &Roberts et al. 1995, ApJ 442, 208;		  \\
tck86:  &Troland et al. 1986, ApJ 304, L57;		  & 
tcgh89: &Troland et al. 1989, ApJ 347, L89;		  \\
tcg+96: &Troland et al. 1996, ApJ 471, 302;		  & 
vg90:   &van der Werf \& Goss, 1990, A\&A 238, 296;	  \\
vgh+93: &van der Werf et al. 1993, ApJ 411, 247.         & \\
\end{tabular}


\end{document}